\documentclass[twocolumn,prd,floats]{revtex4}

\usepackage{graphicx}
\usepackage{amsmath,amsfonts,amssymb}

\begin{document}

\title{Evolution of sub-horizon cold dark matter perturbations}

\author{Christian G. B\"ohmer}
\email{c.boehmer@ucl.ac.uk}
\affiliation{Department of Mathematics and Institute of Origins, University College London, London WC1E 6BT, UK}

\author{Gabriela Caldera-Cabral}
\email{gabycc@math.ucl.ac.uk}
\affiliation{Department of Mathematics and Institute of Origins, University College London, London WC1E 6BT, UK}

\date{\today}

\begin{abstract}
We investigate the evolution of sub-horizon cold dark matter perturbation in the dark energy dominated era of the Universe. By generalizing the Meszaros equation to be valid for an arbitrary equation of state parameter, we derive the $w$-Meszaros equation. Its solutions determine the evolution of the cold dark matter perturbation by neglecting dark energy perturbations. Our analytical results provide a qualitative understanding of this evolution.
\end{abstract}

\maketitle

\section{Introduction}

The cosmological principle is the cornerstone of modern cosmology, it assumes that the Universe is homogeneous and isotropic. The cosmological principle is valid for studying the Universe as a whole but we know that it does not hold perfectly. The nearby Universe is highly inhomogeneous, it is made up of stars, planets and galaxies rather than a smoothly distributed fluid.

One of the merits of inflation is that it provides a theory of inhomogeneities in the Universe, which may explain the observed structures. These inhomogeneities arise from the quantum fluctuations in the inflaton field about its vacuum state, in other words, by the vacuum fluctuation~\cite{Liddle,Dodelson}.

The vacuum fluctuations generate a primeval density perturbation, that is Gaussian and adiabatic, and whose spectral index is close to 1. Such a perturbation was regarded, even before the study of inflation, as a viable candidate for the origin of the large-scale structure and the cosmic microwave background (CMB), which was not been observed at that time~\cite{Peebles}. To understand the evolution of the primeval perturbation to the present, we need to know the nature and amount of the non-baryonic dark matter, as well as the value of the cosmological constant. The simplest possibility is to have zero cosmological constant and cold dark matter giving critical density. That result can fairly be said to be the simplest plausible model for the origin of large-scale structure and the CMB anisotropy and this is called the cold dark matter model (CDM)~\cite{1982ApJ...258..415P,1984Natur.311..517B}.

Setting up a system of equations to be solved and the initial conditions for the perturbations, we can calculate the inhomogeneities and anisotropies in the Universe. The dark matter perturbations are, in principle, coupled to all other perturbations. In practice, however, perturbations to the dark matter depend very little on the details of the radiation perturbations. Dark matter is affected by radiation only indirectly, through the gravitational potentials. At late times, when the Universe is dominated by matter, these potentials are independent of radiation. At early times the gravitational potentials are dominated by radiation. The radiation perturbations are relatively simple, so that all multipole moments beyond the monopole and dipole can be neglected. The converse is not true: to treat the anisotropies properly we will need to know how the matter perturbations behave.   

One of the earliest quantitative investigations of the importance of the gravitational effects of CDM on the baryonic fluctuations which give rise to galaxies was presented by Meszaros in~\cite{1974A&A....37..225M,1975A&A....38....5M,1980ApJ...238..781M}. These papers discussed the evolution of the density contrast in what came to be known as cold dark matter fluctuations in the cosmological radiation background as it enters the horizon and goes through the matter-radiation equilibrium and the recombination epochs, and its effect on the formation of galaxies and large scale structure mediated by such cold dark matter. This is referred to in the literature as the ``Meszaros effect'', which has become incorporated in all subsequent CDM calculations of structure evolution, determining the so-called transfer function of the original fluctuation spectrum. Large scale structure observations, modeling and simulations are now a major activity in cosmology, but the actual nature of dark matter still remains elusive.

In this work, we want to study the evolution of the dark matter perturbations in the dark matter-dark energy era. In Section~\ref{Meszaros_eq} we review the Meszaros equation and its well known solutions. In Section~\ref{wMeszaros} we obtain a Meszaros-like equation: the $w$-Meszaros equation, by using a similar formalism used to obtain the original equation. In Section~\ref{solutions} we find all solutions of this equation and in the final Section~\ref{analysis} we discuss and analyze our results. 

\section{Meszaros equation}
\label{Meszaros_eq}

The radiation pressure causes the gravitational potentials to decay as modes enter the horizon during the radiation era. The pressure suppresses any growth in the radiation perturbations. In contrast, the matter perturbations grow logarithmically. In the radiation era, the potential is determined by radiation but eventually the growth in the matter perturbations offsets that there is more radiation than matter. The dark matter perturbations $(\delta_{c})$ become larger than the radiation perturbations $(\delta_{r})$  even if $\rho_{c}$ is smaller than $\rho_{r}$. Once this happens, the gravitational potential and the dark matter perturbations evolve together and decouple from radiation. 
 
The perturbation equations for dark matter in sub-horizon scales are:
\begin{align}
  \dot \delta_c- v_ck^2 &= 0 
  \label{continuity_gral} \\ 
  -\dot v_c-aHv_ck^2 &= k^2\phi 
  \label{velocity_gral}\\
  -4 \pi G a^2 \rho_c\delta_c &= k^2\phi, \label{poisson_gral}
\end{align}
where the derivative is with respect to the proper time. Note that in the Poisson Eq. (\ref{poisson_gral}) we are not taking into account the radiation perturbations, even when $\phi$ is determined by them, by this time $\delta_{c} >\delta_{r}$ and as we pointed before, $\phi$ and $\delta_c$ evolve together.

We want to follow the perturbations through the epoch of equality, so is better to define a new temporal variable~\cite{Dodelson}:
\begin{align}  
  \frac{\rho_c}{\rho_r}=\frac{a}{a_{\rm eq}} \equiv y,
  \label{defy}
\end{align}
and in terms of this new variable the equations become,
\begin{align}
  \delta '_c-\frac{v_c}{Hay} &= 0\\
  -v'_c-\frac{v_c}{y} &= \frac{\phi}{Hay}\\
  k^2\phi &= -\frac{3y}{2(y+1)}H^2a^2\delta_c,
\end{align}
where the prime denotes differentiation with respect to $y$. By turning this first order system into a second-order differential equation for $\delta_c$ we obtain the Meszaros equation which governs the evolution of sub-horizon cold dark matter perturbations once radiation perturbations have become negligible:
\begin{align}
  \delta''+\frac{2+3y}{2y(y+1)}\delta'-\frac{3}{2y(y+1)}\delta=0. 
  \label{meszaros}
\end{align}
This equation has two solutions given by Meszaros~\cite{1974A&A....37..225M}. The growing solution is,
\begin{align}
  \delta_c^{(1)}=C_1\left(1+\frac{3}{2}y\right),
  \label{growing_mes}
\end{align}
and the decaying solution is given by
\begin{align}
  \delta_c^{(2)}=C_2\left[\left(1+\frac{3}{2}y\right)\ln\frac{\sqrt{1+y}+1}{\sqrt{1+y}-1}-3\sqrt{1+y}\right].
\label{decaying_mes}
\end{align}
Before equality $(y\ll 1)$, the growth mode is practically frozen, whereas at late times $(y \gg 1)$ the growing solution $\delta_c^{(1)}$ scales as $y$ and the solution matches the law in a matter dominated universe
\begin{align}
  \delta_c \propto a.
\end{align}
The decaying mode $\delta_c^{(2)}$ falls off as $y^{-3/2}$.

\section{$w$-Meszaros equation}
\label{wMeszaros}

Now we want to study the evolution of the dark matter perturbations in the dark matter-dark energy era. We are now going to neglect the dark energy perturbations, analogously to neglecting the radiation perturbations in the Meszaros effect previously.

Along the lines of~(\ref{defy}) we introduce a new dimensionless variable as an evolution parameter instead of $t$ or $a(t)$ which we define by
\begin{align}
  y \equiv \left(\frac{a}{a_{\rm co}}\right)^{3w}=\frac{\rho_m}{\rho_x},
\end{align}
where $w$ is the constant equation of state parameter of the dominant component of the Universe, $\rho_m$ corresponds to cold dark matter $\rho_c$ plus baryons $\rho_b$ and $a_{\rm co}$ is the coincidence time: $\left(\frac{\rho_m}{\rho_x}\right)\approx 1$. It should be noted that now $y$ is a decreasing functions with respect to the expansion.

In terms of this variable equations~(\ref{continuity_gral}), (\ref{velocity_gral}) and~(\ref{poisson_gral}) become
\begin{align}
  \delta '_m-\frac{v_m k^2}{3wHay} &=0
  \label{continuity}\\
  -v'_mk^2-\frac{v_m k^2}{3wy} &=\frac{k^2\phi}{3wHay}
  \label{velocity}\\
  -\frac{3y}{2(y+1)}H^2a^2\delta_c &=k^2 \phi
  \label{poisson}
\end{align}
We are going to differentiate Eq.~(\ref{continuity}) in order to turn it into a second-order equation
\begin{align}
  \delta''_m-\frac{v'_mk^2}{3waHy}-v_mk^2\frac{d(3waHy)^{-1}}{dy}=0.
\end{align}
Using the velocity equation Eq.~(\ref{velocity}) we can eliminate $v'_m$ and the continuity equation~(\ref{continuity}) to eliminate the $v_m$-term and from
\begin{align}
  \frac{d}{dy}\left(\frac{1}{aHy}\right)=-\frac{6wy+3w-y-1}{2y^2(y+1)aH}.
\end{align}
the equation becomes
\begin{align}
  \delta''_m+\left[\frac{2+9w^2(3w-1)}{6wy}+\frac{9w^2}{2(1+y)}\right]\delta'_m=-\frac{k^2\phi}{9w^2a^2H^2y^2}
\end{align}
By using the Poisson equation~(\ref{poisson}) we can finally eliminate the gravitational potential $\phi$ and arrive at
\begin{multline}
  \delta''_m+\left[\frac{2+9w^2(3w-1)}{6wy}+\frac{9w^2}{2(1+y)}\right]\delta'_m\\
  -\frac{1}{6w^2y(y+1)}\delta_m=0.
  \label{wmeszaros}
\end{multline}
We refer to this as the $w$-Meszaros equation for dark matter-dark energy, which governs the evolution of sub-horizon cold dark matter perturbations if we neglect dark energy perturbations. This equation tells us that in the dark matter-dark energy era, the dark matter perturbations will decrease as the dark energy starts to dominate. If we set $w=1/3$, for a radiation dominated Universe, we recover the well-known Meszaros equation.

\section{Solving the $w$-Meszaros equation}
\label{solutions}

Let us introduce the new independent variable $z=-y$, then~(\ref{wmeszaros}) can be written in the form
\begin{multline}
  z(1-z)\frac{d^2\delta}{dz^2} + \frac{1}{6w}\Bigl((27w^3 - 9w^2 + 2) \\- (54w^3 -9w^2 + 2) z\Bigr)\frac{d\delta}{dz} + \frac{1}{w^2} \delta = 0,
\end{multline}
which is of the form of the hypergeometric differential equation
\begin{align}
  z(1-z)\frac{d^2\delta}{dz^2} + [c-(a+b+1)z]\frac{d\delta}{dz} - ab\,\delta = 0.
\end{align}
This differential equation has two linearly independent solutions which both involve the hypergeometric function 
\begin{align}
  F[a,b,c;z] = \frac{\Gamma(c)}{\Gamma(a)\Gamma(b)} 
  \sum_{n=0}^{\infty} \frac{\Gamma(a+n)\Gamma(b+n)}{\Gamma(c+n)}\frac{z^n}{n!},
\end{align}
where $\Gamma$ is the Gamma function. The exact form of the solution depends on the numerical values of the constants $a$, $b$, $c$, see~\cite[Chapter 15]{abramowitz+stegun}.

For the $w$-Meszaros equation the three constants $a$, $b$ and $c$ can be expressed in terms of $w$ only, and are given by
\begin{align*}
  a &= \frac{1}{12w}\left(54w^3 - 9w^2 -6w +2 - \sqrt{d} \right)\\
  b &= \frac{1}{12w}\left(54w^3 - 9w^2 -6w +2 + \sqrt{d} \right)\\
  c &= \frac{1}{6w}(27w^3 - 9 w^2 + 2)\\
  d &= 2916 w^6 - 972 w^5 - 567 w^4 + 324 w^3 - 24 w + 28.
\end{align*}
Now, if we choose $w=1/3$ then $a=1$, $b=-3/2$, $c=0$, in which case the general solution involving the hypergeometric function reduces to Eqs.~(\ref{growing_mes}) and~(\ref{decaying_mes}). In principle one can now study solutions of~(\ref{wmeszaros}) numerically for arbitrary $w$. 

However, for some special values of the equation of state parameter, we can find explicit solutions. We found that for the two choices $w=-1/3$ and $w=1/6$ this is possible, and we could not find other values in the parameter range $-1 < w < 1$, including $w=-1$. For $w=-1/3$ the solution is
\begin{multline}
  \delta(y) = c_1 y \sqrt{1 + y} + c_2 \bigl(1 + 3 y \\
  - \frac{3}{2} y \sqrt{1+y} \ln\frac{\sqrt{1+y}+1}{\sqrt{1+y}-1}\bigr),
\end{multline}
which only contains elementary functions. The $c_1$ and $c_2$ part in this solution are the growing and decaying modes with respect to $y$, respectively. This solution is similar to the well-known $w=1/3$ solution, see Eq.~(\ref{growing_mes}) and~(\ref{decaying_mes}).

The solution for $w=1/6$ is given by
\begin{multline}
  \delta(y) = c_1 (3+144y+256 y^2)(1+1/y)^{7/8} \\
  + c_2 (1 + 1/y)^{7/16} Q_2^{7/8}(1+2y),
\end{multline}
where $Q_{\mu}^{\lambda}$ is the generalized Legendre polynomial and we are interested in the real part of this solution. As for the previous solution, the $c_1$ and $c_2$ part are the growing and decaying modes, respectively.

When the equation of state parameter is $w=-2/3$, we have a physically particularly interesting solution since this models one form of dark energy. One cannot write the solution in terms of elementary functions. In terms of the hypergeometric function, we can write
\begin{multline}
  \delta(y) = c_1 \frac{1}{y^{3/2}} F\left[(1-\sqrt{55})/4,(1+\sqrt{55})/4,-1/2,-y\right] \\
  + c_2 F\left[(7-\sqrt{55})/4,(7-\sqrt{55})/4,5/2,-y\right],
  \label{sol23}
\end{multline}
where the constants $c_1$ and $c_2$ are specified by initial or boundary conditions. Note that the independent variable $y$ and the scale factor $a$ are related by $y = 1/a^2$.

\section{Analysis and discussion}
\label{analysis}

\begin{figure*}[!ht]
\centering
\includegraphics[width=0.48\textwidth]{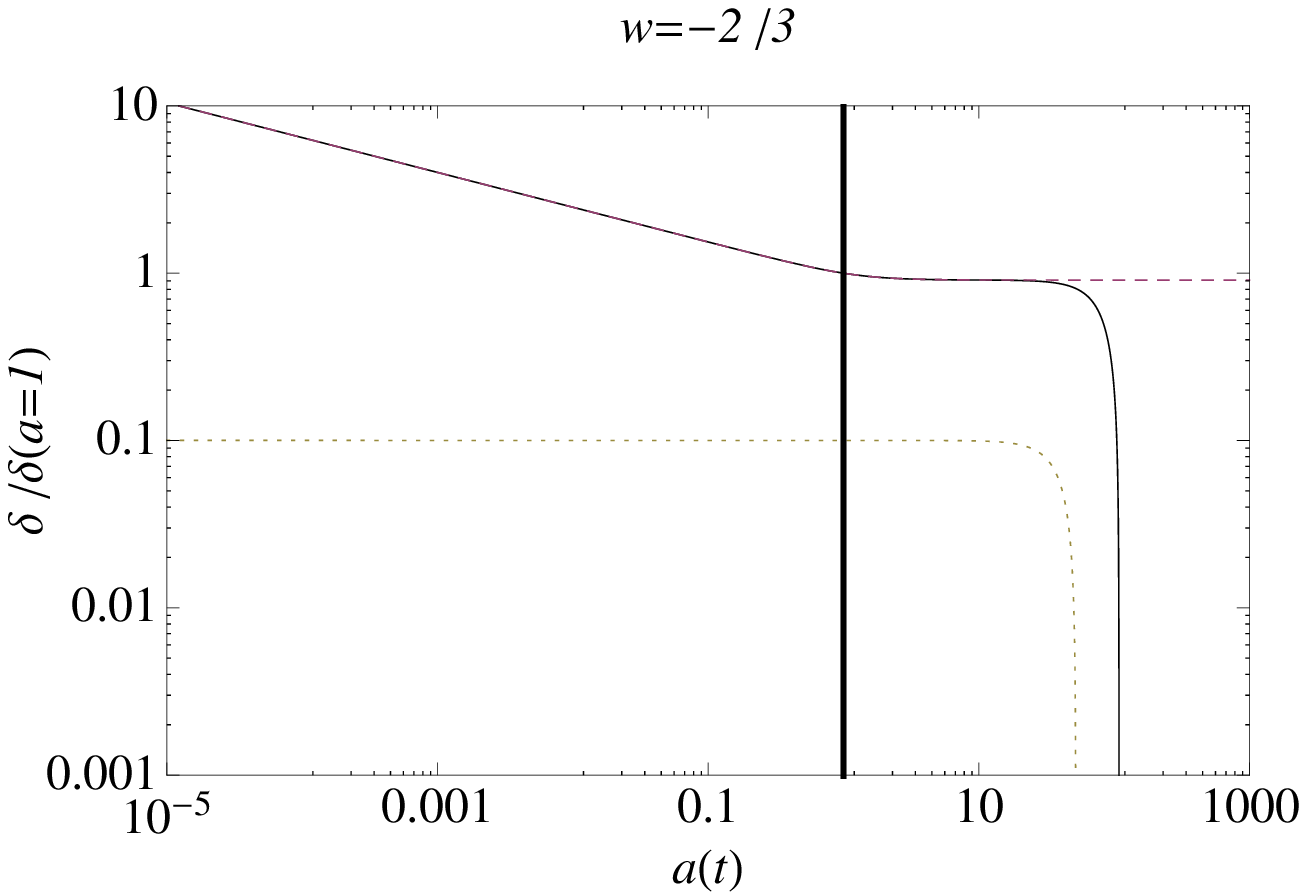}\hfill
\includegraphics[width=0.48\textwidth]{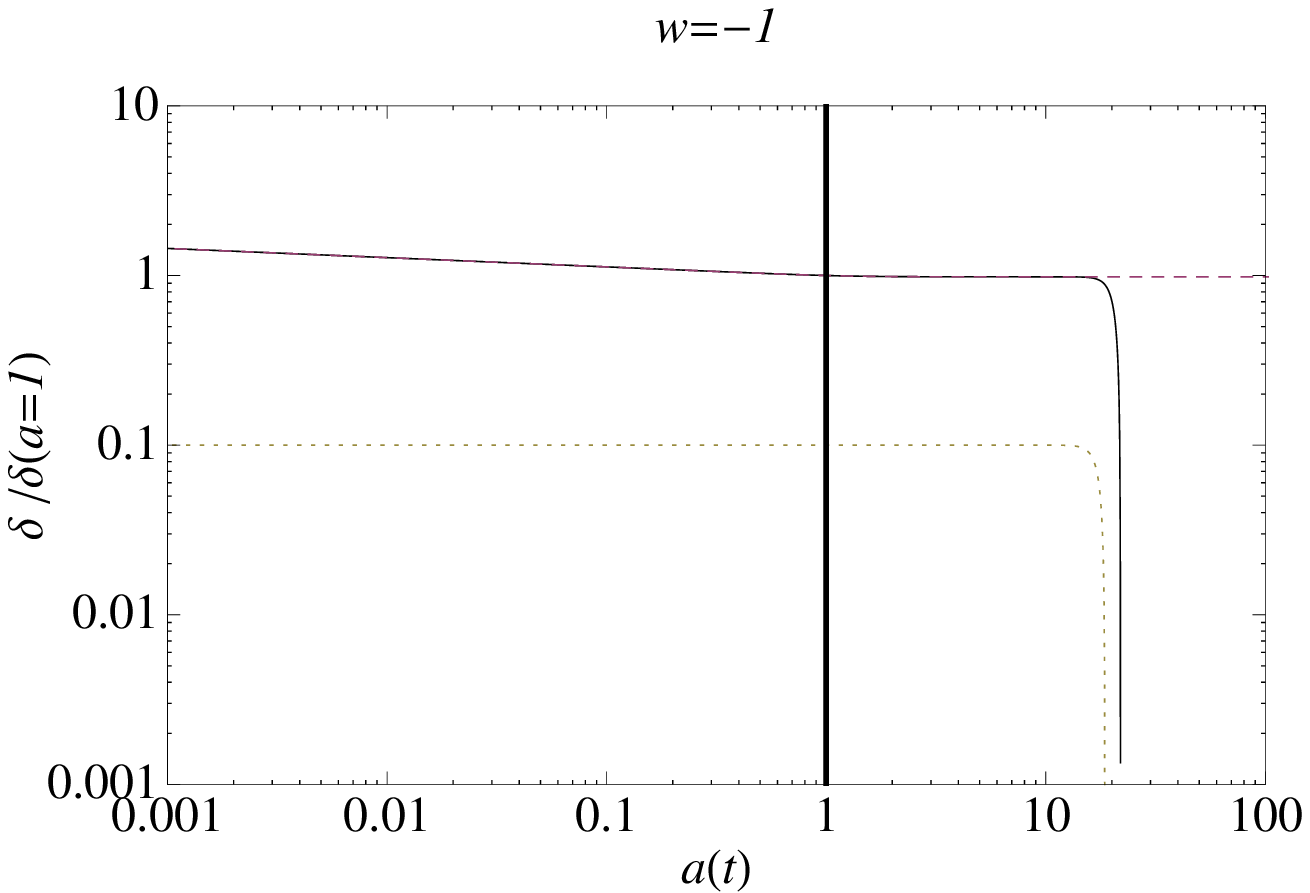}
\caption{The left panel shows the numerical evolution for $w=-2/3$,
the right panel shows the same scenario for $w=-1$. We have chosen the boundary conditions $\delta(y=10^{-4}) = 0.2$ and $\delta(y=1) = 1$. The solid line represents the complete solution, the dotted and dashed lines correspond to the $c_1$ and $c_2$ parts of the solutions respectively. In both plots the dotted lines have been moved to $0.1$ by hand to make them visible.}
\label{mainfig}
\end{figure*}

Despite the fact that solutions of the $w$-Meszaros equation~(\ref{wmeszaros}) are rather complicated, the three constants $a$, $b$ and $c$ depend only on the value of the equation of state parameter $w$. One can easily analyze these solutions numerically and for instance plot the late time evolution of the dark matter perturbations. The left panel of Fig.~\ref{mainfig} shows the evolution of the perturbations $\delta(a)$ of Eq.~(\ref{sol23}) from the dark matter dominated era to the distant future, the dark energy dominated era, for $w=-2/3$. The $a=1$ line corresponds to today. As expected, the dark matter perturbations start to decrease after the time of the coincidence. 

This solution is characterized by two decaying modes which behave rather differently. One part (the $c_2$ part) of solution~(\ref{sol23}) decays for small values of $a$ and then approaches a constant at late times. The other part (the $c_1$ part) of the solution stays constant at early times and rapidly decrease at some point in the near future. This part dominates the late time behavior of the solution which is clearly illustrated in Fig.~\ref{mainfig} (left panel). When we further decrease the value of $w$, we find the same qualitative picture, the right panel in Fig.~\ref{mainfig} shows the solutions for $w=-1$.

As we pointed before the dark matter perturbations start to decrease after the time of the coincidence and will no longer evolve as $\delta_c \approx a$. At some time in the future, when the Universe gets dominated by dark energy they will become negligible and the dark energy perturbations will dominate. This time will depend on the equation of state of the dark energy, i.e.~will depend on the model of dark energy or modified gravity considered to solve the equation. Therefore, the $w$-Meszaros equation can be used to distinguish between different cosmological models. For practical purposes, and to get some predictions out, one would have to use current data and match the solutions in order to find a transfer function by following well-known methods applied to obtain the growth function~\cite{Dodelson}.

It should also be emphasized that one can, in principle, use the $w$-Meszaros equation to study phantom dark energy models, where $w < -1$, see for example~\cite{Caldwell:2003vq,2006IJMPD..15.1299A,PhysRevD.71.047301,Nesseris:2006er}. The treatment of perturbations in such models can be quite involved, however, our equations are well defined for all $w$. We suggest to use this equation to gain a preliminary understanding of phantom dark energy perturbations in the late time universe.

Another interesting idea we would like to point out is the derivation of a Meszaros like equation in situations where couplings between dark matter and dark energy are allowed. This would, as before, allow for studying a large class of models qualitatively, see also~\cite{CalderaCabral:2009ja}.

Finally, solving the $w$-Meszaros equation can give us an idea of when the dark energy perturbations become important and at what time (size of the universe) they have to be taken into account in calculations for an specific model. Note that in this work we are not considering models with early time dark energy that can lead to some early time presence of dark energy perturbations. Therefore, it would be very interesting to study the evolution of dark energy perturbations using the method described here, however, this is beyond the scope of this brief report.

\acknowledgments
We would like to thank Roy Maartens, Rod Halburd and Yuri Obukhov for useful discussions on the manuscript. GC-C is supported by the Mexican Council of Science and Technology (CONACyT), postdoctoral fellowship grant 120144.

\bibliographystyle{apsrev}
\bibliography{meszaros}

\end{document}